\documentclass[conference]{IEEEtran}
\IEEEoverridecommandlockouts
\usepackage{cite}
\usepackage{amsmath,amssymb,amsfonts}
\usepackage{algorithmic}
\usepackage{graphicx}
\usepackage{textcomp}
\usepackage{xcolor}
\usepackage{booktabs}
\usepackage{tabularx}
\usepackage{algorithm}
\usepackage{pgfplots}
\usepackage{balance}
\pgfplotsset{compat=1.18}

\newtheorem{definition}{Definition}
\newtheorem{theorem}{Theorem}
\newtheorem{problem}{Problem}
\newtheorem{proof}{Proof}
\newtheorem{proposition}{Proposition}
\def\BibTeX{{\rm B\kern-.05em{\sc i\kern-.025em b}\kern-.08em
    T\kern-.1667em\lower.7ex\hbox{E}\kern-.125emX}}

\begin{document}

\title{Automatic Generation of Safety-compliant Linear Temporal Logic via Large Language Model: A Self-supervised Framework\\

}

\author{
    \IEEEauthorblockN{
        Junle Li\textsuperscript{1}, 
        Siqi Chen\textsuperscript{1}, 
        Jiakai Li\textsuperscript{1}, 
        Meiqi Tian\textsuperscript{1}, 
        Bingzhuo Zhong\textsuperscript{1*}
    }
    \IEEEauthorblockA{
        \textsuperscript{1}\textit{Hong Kong University of Science and Technology (Guangzhou)}, Guangzhou, China \\
        Email: \{juliejunleli, bingzhuoz\}@hkust-gz.edu.cn, \{schen972, jli196, mtian837\}@connect.hkust-gz.edu.cn
    }

    \thanks{*Corresponding author.}
}

\maketitle

\begin{abstract}
Converting high‑level natural‑language task descriptions into formal specifications such as Linear Temporal Logic (LTL) is essential for ensuring safety in cyber‑physical systems (CPS). Existing work, however, \emph{only} optimizes translation quality without explicitly verifying the output against safety constraints. We present \texttt{AutoSafeLTL}, a self‑supervised, cloud–edge–collaborative framework that automates the generation of safety‑compliant LTL specifications while preserving logical consistency and semantic fidelity. A lightweight edge‑side three-stage-fine-tuned LLM offers real‑time conversion from natural language to LTL specifications (NL2LTL) and \emph{guarantees} safety‑critical latency and data locality. Two larger‑capacity cloud‑side agents then iteratively refine the alignment: 1) \emph{LLM‑as‑an‑Aligner} matches atomic propositions to safety constraints, and 2) \emph{LLM‑as‑a‑Critic} interprets counterexamples from Inclusion Check to guide corrective regeneration. This collaborative architecture provides a safety-guaranteed alignment mechanism between high-level user intent and formally verifiable system behavior, demonstrating the potential of our framework to advance AI Alignment in safety-critical domains. Our approach achieves 0\% violation rates on multiple benchmarks, enabling trustworthy specification generation and verification for both AI and critical CPS applications.
\end{abstract}


\section{Introduction}
\textbf{Motivation:}
Ensuring the safety of cyber-physical systems (CPS) is central to modern system design, especially in safety-critical domains. Formal specifications like Linear Temporal Logic (LTL) \cite{43} provide mathematical rigor for verification and controller synthesis, ensuring CPS safety. The challenge lies in translating high-level, often unstructured requirements into formal specifications.
However, a critical issue frequently neglected by these methods is whether the generated LTL formula inherently conflict with the safety constraints imposed on the system. We refer to this aspect as the \emph{compliance of LTL specifications} with respect to safety constraints, and to those LTL specifications with no such conflict as \emph{safety-compliant LTL}. Specifically, being safety-compliant means that the temporal behavior expressed by the LTL formula entirely aligns with the intended temporal behavior mandated by safety constraints. Accordingly, we focus on how to generate \emph{safety-compliant LTL} specifications via Large Language Models (LLMs), while maintaining formal safety-compliance guarantees to prevent critical risks to CPS.

\textbf{Related Works:}
Linear Temporal Logic (LTL) is fundamental for formal verification in autonomous systems. Recent NL2LTL approaches improve translation efficiency and syntactic validity via decomposition, dynamic prompting, grammar-constrained decoding, and fine-tuning \cite{14}. However, these methods focus exclusively on \emph{semantic alignment}. By treating user instructions as the absolute truth, they ignore \emph{external safety constraints}, risking the faithful translation of unsafe intents into dangerous specifications. Meanwhile, frameworks integrating LLMs with formal verifiers typically position the verifier as a \emph{post-hoc filter} for LLM-generated plans. This remedial paradigm leads to inefficient ``trial-and-error'' cycles and only verifies one-off execution traces rather than the underlying task logic, lacking a persistent safety foundation against environmental variations.

\begin{figure}[t]  
    \centering
    \includegraphics[width=\columnwidth]{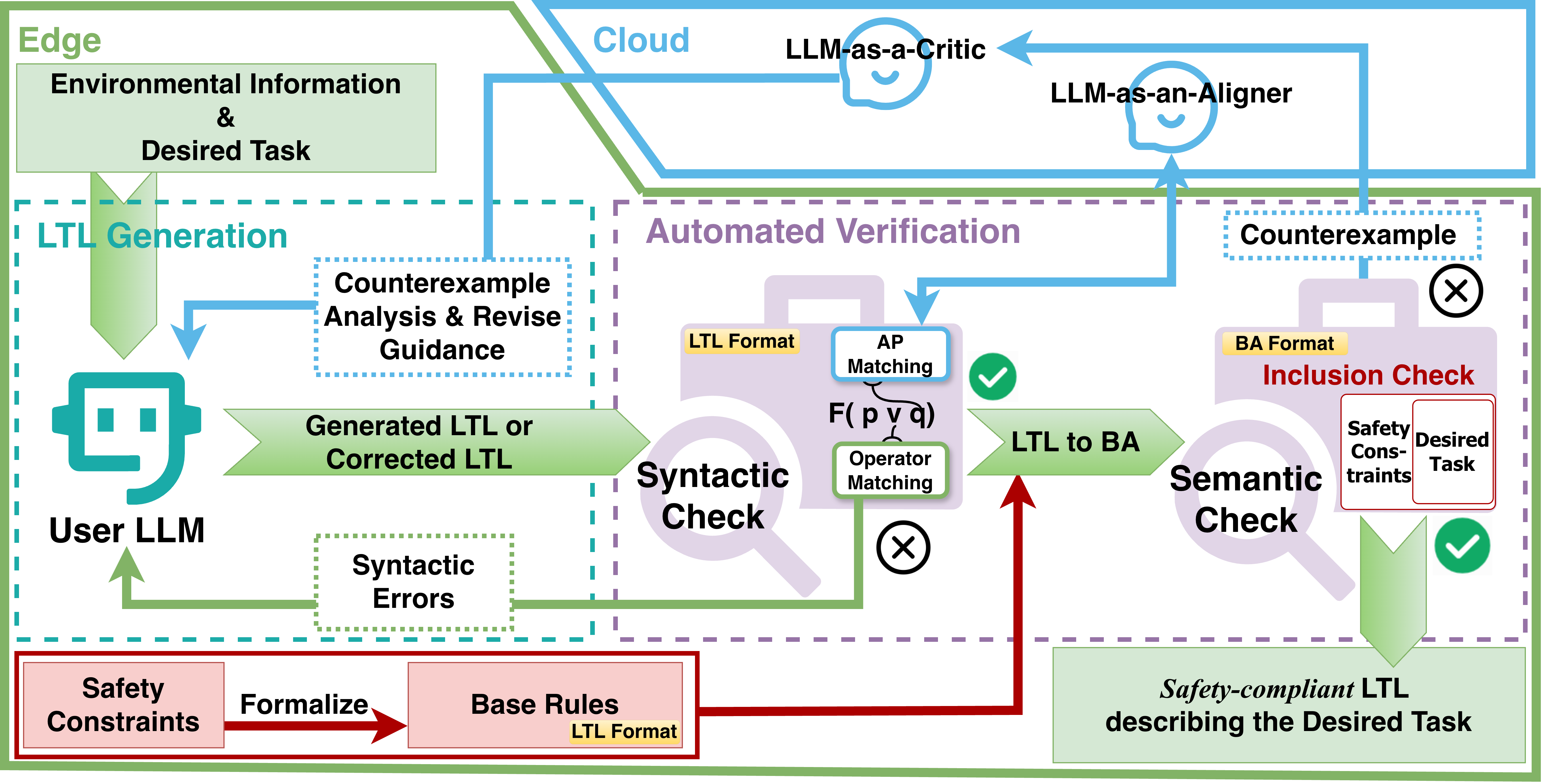} 
    \vspace{-0.3cm}
    \caption{Overview of \texttt{AutoSafeLTL} Framework.}
    \label{fig:Frame}
    \vspace{-0.2cm} 
\end{figure}

In this work, we present \texttt{AutoSafeLTL} (as shown in Fig. \ref{fig:Frame}), a self-supervised framework that integrates automata-based verification with a cloud–edge collaborative LLM architecture to generate \emph{safety-compliant} LTL specifications in a fully-automated manner. An edge-deployed \emph{User LLM}, fine-tuned in three progressive stages (translation, syntactic, and semantic correction), handles both real-time LTL generation and local repair with minimal latency. For rigorous verification, cloud-side agents—\emph{LLM-as-an-Aligner} and \emph{LLM-as-a-Critic}—are invoked to interpret complex violations detected via Inclusion Checks. This collaborative synergy ensures data locality while leveraging cloud-scale inference to achieve formal alignment between user intent and safety properties. Our contributions are summarized as follows:
\vspace{-0.05cm}
\begin{itemize}
    \item We propose a \emph{fully automated}, end-to-end framework for NL-driven LTL specification generation that targets \emph{early-stage safety-compliant} specification generation. By shifting the focus to the generation of the specification itself, our framework ensures the governing logic is formally verified before system execution. We enhance formal verification tools with an efficient counterexample path extraction algorithm, enabling \emph{automated} LTL refinement and a 0\% violation rate against safety constraints.
    
    \item We develop a novel three-stage fine-tuning strategy that progressively instills \emph{intrinsic safety} capability into a small language model. This strategy enables an 8B edge model to outperform GPT-4 in generating safety-compliant LTL specifications, proving the viability of formal methods on resource-restricted  hardware.

    \item We establish a cloud–edge collaborative framework that enforces data locality through on-device LTL generation and correction. By delegating abstracted tasks—AP alignment and counterexample interpretation—to cloud-side agents, this architecture enables the scalable deployment of formal methods in safety-critical and privacy-sensitive domains.

    \item We construct fine-tuning and evaluation datasets to address the lack of suitable datasets for evaluating safety-compliance in NL2LTL tasks. Specifically, our datasets comprise over 35k synthetic LTL specifications for pre-training, 200 carefully curated human–LTL pairs for fine-tuning to enhance the model’s correction ability, and 300 evaluation cases, featuring navigation-style instructions with human-centered APs and rich temporal structures, enabling safety-realistic evaluation.
\end{itemize}
\vspace{-0.1cm}

\section{Preliminaries}
This section introduces the necessary background for our framework. We begin with an overview of \emph{Linear Temporal Logic} (LTL), then introduce \emph{Büchi automata}, which serves as the automata-theoretic counterpart to LTL formulas. Building upon this foundation, we describe the process of \emph{automata-based language inclusion check} to reason the verification. Finally, we formally state the problem that our work aims to address.

\vspace{-0.2cm}
\subsection{Linear Temporal Logic}\label{sec:LTL}
Linear Temporal Logic (LTL) serves as the foundation for many practical specification languages \cite{43}. The basic components of an LTL formula include Atomic Propositions (AP) and Operators (boolean and temporal). 
In our work, we adopt the syntax and semantics of LTL as defined in \cite{23,28} and recalled them here.

\begin{definition}[LTL Syntax and Semantics]
\noindent\textbf{Syntax.}  
Let $AP$ be a finite set of atomic propositions. The set of LTL formulas over $AP$ is defined inductively by the following grammar:
\vspace{-0.1cm}
\[
    \varphi ::= \text{true} \mid a \mid \neg \varphi \mid \varphi_1 \land \varphi_2 \mid X \varphi \mid \varphi_1 \, U \, \varphi_2,
\]
where $a \in AP$ is an atomic proposition, temporal operators such as $X$ (next), $U$ (until) can be used for dynamical properties. 

\noindent\textbf{Semantics.}
Let $\varphi$ be a LTL formula, the semantics of  $\varphi$ is defined over infinite words $\sigma \in (2^{AP})^{\omega}$, with $Words(\varphi)=\left\{\sigma \in\left(2^{AP}\right)^{\omega} \mid \sigma \models \varphi\right\}$ being the LT property induced by $\varphi$. Let $\sigma = A_0 A_1 A_2 \ldots$ (symbol $A_{i}\in AP$), where the satisfaction relation $\models \subseteq (2^{AP})^{\omega} \times \text{LTL}$ is defined as follows:
\vspace{-0.45cm}
\begin{itemize}
    \item $\sigma \models \text{true}$,
    \item $\sigma \models a$ iff $a \in A_0$,
    \item $\sigma \models \neg \varphi$ iff $\sigma \not\models \varphi$,
    \item $\sigma \models \varphi_1 \land \varphi_2$ iff $\sigma \models \varphi_1$ and $\sigma \models \varphi_2$,
    \item $\sigma \models X \varphi$ iff $\sigma[1...] = A_1 A_2 \ldots \models \varphi$,
    \item $\sigma \models \varphi_1 \, U \, \varphi_2$ iff $\exists j \geq 0$. $\sigma[j...] \models \varphi_2$ and for all $0 \leq i < j$, $\sigma[i...] \models \varphi_1$.
\end{itemize}
\label{def:LTL}
\end{definition}

\vspace{-0.2cm}
\subsection{Büchi automaton}\label{subsec:BA}
To bridge the gap between logical formulations and algorithmic reasoning, a typical step is to convert LTL into automata, like Büchi automaton (BA) \cite{2}.

\begin{definition}[Büchi Automata]
    A Büchi automaton (BA) is defined as a tuple $\mathcal{A} = (\Sigma, Q,\mathcal{I}, \mathcal{F},\delta)$, where $\Sigma$ is an finite alphabet,  $Q$ is a finite set of states, $\mathcal{I} \subseteq Q$ is the set of initial states, $\mathcal{F} \subseteq Q$ is the set of accepting states and $\delta: Q \times \Sigma \times 2^Q$ is the transition relation. A run of $\mathcal{A}$ on an infinite word $w = \sigma_1 \sigma_2 \dots \in \Sigma^\omega$, with symbol $\sigma_i \in \Sigma$, $i\in \mathbb{N}$, starting from a state $q_0 \in Q$, is an infinite sequence of states $q_0 q_1 q_2 \dots$ such that $(q_{j-1}, \sigma_j, q_j) \in \delta$ for all $j > 0$. The run is accepted if $q_i \in F$ for infinitely many $i$. The accepted language $L(\mathcal{A})$ consists of all infinite words over $\Sigma^\omega$ for which there exists an accepted run.
    \end{definition}

For every LTL formula $\varphi$, there exists a corresponding Büchi automaton $\mathcal{A}$ such that:
\(
L(\mathcal{A})=Words(\varphi)
\). The conversion from LTL to Büchi automaton has a well-established theoretical foundation \cite{25,26}, and existing research has proposed numerous effective methods and tools for converting LTL to Büchi automata \cite{27,28,29,30,31,32,33}. The aim of this conversion lies in using Büchi automata to model a system’s dynamic behavior. Simulation relations, such as forward and backward simulation, are then used to establish step-by-step correspondence by linking successive state transitions, thereby capturing all dynamic behaviors expressed by the BA.

\begin{definition}[Forward and Backward Simulation on Büchi Automata]
Let $\mathcal{A} = (\Sigma, Q,\mathcal{I}, \mathcal{F},\delta)$ be a Büchi automaton. A relation $R \subseteq Q \times Q$ is called (1) a \emph{forward simulation} on $\mathcal{A}$ if for any $(p, r) \in R$, the following conditions hold: 1)If $p \in \mathcal{F}$, then $r \in \mathcal{F}$ and 2) $\forall(p, \sigma, p') \in \delta$, $\exists(r, \sigma, r') \in \delta$ s.t. $(p', r') \in R$. (2) a \emph{backward simulation} on $\mathcal{A}$ if for any $(p', r') \in R$, the following conditions hold: 1) If $p' \in \mathcal{F}$, then $r' \in \mathcal{F}$, 2) If $p' \in \mathcal{I}$, then $r' \in \mathcal{I}$, and 3) $\forall (p, \sigma, p') \in \delta$, $\exists(r, \sigma, r') \in \delta$ s.t. $(p, r) \in R$.
\end{definition}

To support the subsequent Section~\ref{section3.3}, we introduce the notion of a \emph{path} in a Büchi automaton, which uses simulation to describe how an input word drives a sequence of state transitions. Since the same symbol \( \sigma_i \in \sigma \) can trigger different transitions depending on the current state of the automata, there exists a set of paths over the same input word $\sigma$.

\begin{definition}[Path of a Büchi Automaton]
Let $\mathcal{A} = (\Sigma, Q,\mathcal{I}, \mathcal{F},\delta)$ be a Büchi automaton and let Word $\sigma = \sigma_0 \sigma_1 \sigma_2 \ldots \in \Sigma^{w}$ 
 with symbol $\sigma_i \in \Sigma$, $i\in\mathbb{N}$. A \emph{path} $\rho$ of $\mathcal{A}$ over $\sigma$ is a state-and-word sequence:
\vspace{-0.1cm}
\[
q_0 \xrightarrow{\sigma_0} q_1 \xrightarrow{\sigma_1} q_2 \xrightarrow{\sigma_2} \cdots
\]
where $q_0 \in \mathcal{I}$ is an initial state and for each $i \geq 0$, $(q_i, \sigma_i, q_{i+1}) \in \delta$. 
A path $\rho$ is \emph{simulable} w.r.t. $\sigma$ if: 1) it traverses every $\sigma_i \in \sigma$  and the corresponding transitions $(q_i, \sigma_i, q_{i+1}) \in \delta$;  
2) it ends in or visits some accepting state $q_f \in \mathcal{F}$ infinitely often. The set of paths over $\sigma$ in \( \mathcal{A} \) is denoted by \( \mathcal{P}_\sigma \), where:
\vspace{-0.1cm}
\begin{equation}\label{Psigma}
    \mathcal{P}_\sigma = \{ q_0 \xrightarrow{\sigma_0} q_1 \xrightarrow{\sigma_1} \cdots \mid q_i \in Q,\ \sigma_i \in \sigma,\ (q_i, \sigma_i, q_{i+1}) \in \delta \} 
\end{equation}
\label{def:Path} 
\end{definition}
\vspace{-0.4cm}

A fundamental concept in automata-based verifications is the notion of \emph{language inclusion}, capturing the containment relationship between two automata.

\vspace{-0.1cm}
\begin{definition}[Language Inclusion]
Let $\mathcal{A}_1$ and $\mathcal{A}_2$ be two Büchi automata over the same input alphabet $\Sigma$. If every infinite word accepted by $\mathcal{A}_1$ is also accepted by $\mathcal{A}_2$, then the language of $\mathcal{A}_1$ is \emph{included} in the language of $\mathcal{A}_2$, denoted as $L(\mathcal{A}_1) \subseteq L(\mathcal{A}_2)$:
\vspace{-0.1cm}
\[
L(\mathcal{A}_1) \subseteq L(\mathcal{A}_2) \iff \forall w \in \Sigma^\omega, \; w \in L(\mathcal{A}_1) \Rightarrow w \in L(\mathcal{A}_2).
\]
\end{definition}
\vspace{-0.3cm}

\subsection{Automata-Based Language Inclusion Checking among LTLs}

The results in \cite{esterle2020formalizing} shows that safety restrictions could also be interpreted as LTL (therefore BA).
Therefore, the compliance of LTL specifications can be converted to an Automata-based language inclusion check \cite{57}, with rank-based and Ramsey-based methods being the most prominent approaches \cite{35,36,37,38,39,40}. 
Typically, the Ramsey-based method is rooted in the concept of \textit{supergraph}—a data structure representing finite word classes with similar behavior between two automata, to capture the violation counterexample \cite{40}.

\begin{definition}[Supergraph in Ramsey-Based Inclusion Checking]
Let \( \mathcal{A}_1 = (\Sigma, Q_1,\mathcal{I}_1, \mathcal{F}_1,\delta_1) \) and \( \mathcal{A}_2 =  (\Sigma, Q_2,\mathcal{I}_2, \mathcal{F}_2,\delta_2) \) be two Büchi automata over the same alphabet \( \Sigma \). A \emph{supergraph} \( \mathcal{G}( V, E) \) is a finite directed graph constructed to capture the combined behaviors of \( \mathcal{A}_1 \) and \( \mathcal{A}_2 \). The set of nodes \( V \subseteq Q_1 \times Q_2 \) consists of all pairs of states \( (q_1, q_2) \) where \( q_1 \in Q_1 \) and \( q_2 \in Q_2 \), the set of edges \( E \subseteq V \times \Sigma \times V \) captures synchronized transitions under the same input symbol \( a \in \Sigma \), s.t.:
\(
\text{If } (q_1, a,q_1' )\in \delta_1 \text{ and } (q_2, a,q_2' )\in \delta_2, \text{ then } ((q_1, q_2), a, (q_1', q_2')) \in E.
\)
\label{def:supergraph}
\end{definition}

\subsection{Problem Formulation}
\vspace{-0.1cm}
\begin{problem}[Safety-Compliant LTL Specification Synthesis]
Let \( \mathcal{D} \) be an unstructured natural language description of a desired task, and let $\Psi = \{\psi_1, \psi_2, \dots,$ $ \psi_n\} $ be a set of predefined safety restrictions, each expressed as an LTL formula. Design a framework that synthesizes an LTL formula \( \varphi \) based on \( \mathcal{D} \), such that: 
\(
L(\mathcal{A}_\varphi) \subseteq \bigcap_{i=1}^{n} L(\mathcal{A}_{\psi_i}),
\)
where \( \mathcal{A}_\varphi \) and \( \mathcal{A}_{\psi_i} \) are the Büchi automata corresponding to \( \varphi \) and \( \psi_i \), respectively. 
\end{problem}

\section{Framework}\label{sec:Auto}
\vspace{-0.2cm}
\subsection{Overview}

We propose \texttt{AutoSafeLTL} (Fig.~\ref{fig:Frame}), comprising two collaborative modules: \emph{LTL Generation} and \emph{Automated Verification}. Given \emph{Desired Tasks} and \emph{Environmental Information} in natural language, a lightweight edge-side \emph{User LLM}—fine-tuned in three stages —generates and locally revises candidate LTL specifications.

The Automated Verification process consists of two stages: \emph{Syntactic Check} and \emph{Semantic Check}. 
Syntactic Check ensures grammatical validity.
After passing the Syntactic Check, the  Semantic Check ensures that the properties of state paths expressed by the candidate LTL specification adhere to safety constraints.
In particular, both the Desired Task in LTL-format and the formalized \emph{safety constraints} are converted into BA format for the Inclusion Check. 

The principles of feedback and regeneration further ensure safety-compliance. Two cloud-side \emph{Agent LLMs} iteratively refine the LTL specifications: 1) \emph{LLM-as-an-Aligner} ensures atomic proposition alignment; and 2) \emph{LLM-as-a-Critic} guides corrective regeneration using Language Inclusion counterexamples. This cloud-edge synergy balances data locality and rigorous safety verification, highlighting a practical path towards AI alignment for safety-critical CPS applications. For a better illustration of the framework, we used the following
running example throughout the entire session.

\textbf{Running example}
Consider a traffic scenario, an NL navigation task along with a related safety constraint. Our goal is to automatically generate an LTL specification that represents the desired task while adhering to the safety constraint. This specification can then support downstream applications such as path planning.

\textbf{Desired Task:} Start from the current lane, go straight for 200m. Then, turn right onto Maple St. and proceed for 500m. Finally, turn left onto Oak St., and you will arrive at the destination after 300m. 

\textbf{Env. Info:} Elm St. (Straight lane, 50 km/h limit); Car overtaking on right; Target: Oak St. (1 km away).

\textbf{Safety Constraint:} $G((\text{straight\_500m} \vee \text{right\_turn}) \rightarrow (\text{right\_turn} \rightarrow \text{straight\_500m} \rightarrow \text{left\_turn})) \rightarrow G(\text{straight\_1km} \rightarrow \text{arrive\_destination})$. \hfill *














\textit{Remark 1.} While the running example and the evaluation part in Section~\ref{sec:eva} focus on a traffic scenario for illustration, our framework allows for flexible addition and adaptation of desired tasks and specifications as needed.\\
\textit{Remark 2.} While Environmental Information is not required to generate the initial LTL formula, it is essential for LLMs to \emph{regenerate} safety-compliant specifications when violations occur, especially when the original task description is inherently unsafe. 
\vspace{-0.2cm}
\subsection{LTL Generation}\label{sec:ltl-Generation}
The LTL Generation module performs the initial NL2LTL translation from task descriptions and auxiliary environmental information on the edge. We deploy a lightweight \emph{User LLM} that has been sequentially fine-tuned in three stages—NL2LTL translation (S1), syntactic correction  (S2), and semantic correction (S3)—so that a single on-device model jointly handles generation and subsequent local repair, satisfying the CPS-driven requirement: data locality for privacy, or regulation-sensitive deployments.

Given an NL input \( \mathcal{D} \)  and context $E$, \emph{User LLM} extracts an initial LTL $\varphi^{(0)}$. This output is then passed to the Automated Verification module. Fine-tuning details are illustrated in Section \ref{sec:4}.
\vspace{-0.2cm}
\subsection{Automated Verification}\label{section3.3}
Verification ensures that the generated LTL specification satisfies safety constraints. 
Our Automated Verification scheme, the core of the framework, comprises a \emph{Syntactic Check} and a \emph{Semantic Check}.

\subsubsection{Syntactic Check}
The syntactic check consists of two steps: Atomic Proposition (AP) Matching and Operator Matching.
For AP Matching, APs are extracted from safety constraints to build a library. The cloud-side \emph{LLM-as-an-Aligner} aligns APs in the Desired Task with entries in this library and replaces them with the closest matches to ensure AP consistency.
For Operator Matching, a token-based algorithm checks operator usage and parenthesis balance.
Any detected error (along with its type and position) is fed to \emph{User LLM} for immediate on-device correction.

Now, we revisit the running example to demonstrate the syntactic check.\\
\textbf{Example (continued)}
Having the safety constraint in LTL-format, we extract the AP library as \begin{small}
    [\texttt{right\_turn}, \texttt{straight\_500m},\texttt{left\_turn},\texttt{straight\_1km}, \texttt{arrive\_destination}].\end{small}
Then, \emph{LLM-as-an-Aligner} is invoked to compare the LTL-formatted Desired Tasks with the AP Library. 
Accordingly, 
\begin{small}    \texttt{go\_straight\_500m} \end{small}in the original LTL formula is replaced with \begin{small}\texttt{straight\_500m}\end{small} from the AP Library due to their similarity, and \begin{small}\texttt{go\_straight\_200m}\end{small} is replaced with the format-similar AP \begin{small}\texttt{straight\_200m}\end{small}.
All other parts in the LTL formula remain unchanged, which results in a new LTL formula as follows:\\
\vspace{-0.3cm}
\begin{small}
    \begin{align}
    & F(\text{straight\_200m}) \land X \Big( G \Big( \text{right\_turn\_Maple\_St} \to \nonumber \\
    & F\big(\text{straight\_500m} \land X(\text{left\_turn\_Oak\_St} \land F(\text{straight\_300m}))\big) \Big) \Big)
    \label{eq:ltl_corrected}
\end{align}
\end{small}

In the Operator Matching step, both are correctly matched. 
The resulting LTL then undergoes the Inclusion Check. \hfill *

\subsubsection{Inclusion Check}
After the syntactic check, the LTL formulas for the Desired Task and safety constraints are converted into BA format \texttt{Input\_BA} and  \texttt{Comparison\_BA} for the Inclusion Check. We adopt \texttt{Spot} \cite{33} for LTL2BA translation and \texttt{RABIT} \cite{37} for Inclusion Check. Rather than simply integrating these formal tools, we adapt and enhance them to meet specific application objectives, while keeping the pipeline compatible with alternative solutions.\\
\textit{Remark 3.} In practice, multiple safety constraints may exist as separate LTL formulas. Our framework verifies each formula in parallel, supports adding new rules dynamically, and allows prioritizing checks by importance to provide different levels of safety assurance.
To provide more intuition for the Inclusion Check, we now revisit the running example.\\
\textbf{Example (continued)} Both the Safety Constraint and Desired Task are converted into Büchi automata and then subjected to a Ramsey-based Inclusion Check. The process terminates once a counterexample is found, which in this case is \texttt{!straight\_200m}. This counterexample guides the subsequent LTL correction. \hfill *







\subsubsection{Counterexample-guided Automatic Correction}\label{subsubect:regene}
In Language Inclusion Check, a single counterexample word $\sigma$ can be found if there exist any violations, but $\sigma$ may correspond to multiple dynamic behaviors, as the same symbol \( \sigma_i \in \sigma \) can trigger different transitions depending on the current state of the automata. Consequently, identifying a concrete counterexample path - a sequence of state transitions combined with the driving symbol - is essential for accurately locating the source of the violation. This path serves as a precise semantic witness for non-inclusion and provides actionable guidance for correcting the LTL specification. To this end, we formally define the notion of a counterexample path (cf. Section~\ref{subsec:BA}: Definition~\ref{def:Path}) and establish its existence Theorem~\ref{theo:counterexample}. Based on that, we integrate path extraction into the Inclusion Checking process via the Comparison Path Storage Algorithm~\ref{alg:TraceStore}. The worst case complexity of extracting a counterexample path is $O(n_1 \times n_2 \times m)$, where $n_1$ and $n_2$ are the state sizes of the two Büchi automata and $m$ is the alphabet size. The formal proposition \ref{prop:time} and its proof \ref{proof:time} are provided. Together, this formal theory establishes a rigorous and actionable foundation for counterexample-guided LTL correction, constituting a key theoretical contribution of our work.

\begin{theorem}[Existence of Counterexample Path]
Let $\mathcal{A}_1 = (\Sigma, Q_1,\mathcal{I}_1, \mathcal{F}_1,$ $\delta_1)$ and \( \mathcal{A}_2 = (\Sigma, Q_2,\mathcal{I}_2, \mathcal{F}_2,\delta_2) \) be two Büchi automata. 
If there exists an infinite counterexample word \(\sigma= \sigma_1(\sigma_2)^\omega \in \Sigma^\omega \), where \( \sigma_1 \in \Sigma^* \) is a finite prefix, and \( \sigma_2 \in \Sigma^+ \) is a non-empty finite loop, such that
$\sigma$ is accepted by $\mathcal{A}_1$ but rejected by $\mathcal{A}_2$ (i.e., witnessing \( L(\mathcal{A}_1) \not\subseteq L(\mathcal{A}_2) \)),  then a finite counterexample path $\rho \in \mathcal{P}_\sigma$ on $\sigma$ exists, with $\mathcal{P}_\sigma$ as in~\eqref{Psigma}, which is simulable in $\mathcal{A}_1$ but not simulable in $\mathcal{A}_2$. 
\label{theo:counterexample}
\end{theorem}

\begin{proof}  
Let \( \sigma = \sigma_1(\sigma_2)^\omega \in L(\mathcal{A}_1) \setminus L(\mathcal{A}_2) \) be a counterexample word. By definition, there exists an accepting path of \( \mathcal{A}_1 \) on \( \sigma \):  
\vspace{-0.3cm}

\[
\rho_1: q_{1,0} \xrightarrow{\sigma_1} q_{1,c} \xrightarrow{\sigma_2} q_{1,c} \xrightarrow{\sigma_2} \cdots,
\]

where \( \sigma_{2} \) visits accepting state(s) infinitely. 
Assuming that there is no finite counterexample path $\rho$ on word $\sigma$ that is simulable in $\mathcal{A}_1$ but not simulable in $\mathcal{A}_2$ exists. 
Then, for all \( n \in \mathbb{N} \), the finite path \( \rho^{(n)} = \sigma_1(\sigma_2)^n \) must be simulable by \( \mathcal{A}_2 \); that is, \( \mathcal{A}_2 \) can simulate a path: 
\vspace{-0.2cm}

\begin{equation*}
    \rho_{2}: q_{2,0} \xrightarrow{\sigma_1} q_{2,c}^{(n)} \xrightarrow{\sigma_2^n} \tilde{q}_{2,c}^{(n)}, \text{with }  q_{2,c}^{(n)} ,\tilde{q}_{2,c}^{(n)} \in Q_2,
\end{equation*}

Since \( \sigma \notin L(\mathcal{A}_2) \), every infinite path of \( \mathcal{A}_2 \) on \( \sigma \) either:  
1) Fails to being driven by \( \sigma_1 \) (terminating at some \( i \leq |\sigma_1| \)), or  
2) Processes \( \sigma_1(\sigma_2)^\omega \) but fails to satisfy the Büchi condition (an $\omega$-regular acceptance condition) (Büchi,1996). 

Case 1 (Finite Rejection): 
Suppose \( \mathcal{A}_2 \) cannot be driven by \( \sigma_1 \) fully. Let \( i \leq |\sigma_1| \) be the minimal position where no transition in \( \mathcal{A}_2 \) matches \( \mathcal{A}_1 \)’s step \begin{small}\( q_{1,i} \xrightarrow{\sigma_{i+1}} q_{1,i+1} \)\end{small}. The finite path \begin{small}$\quad q_{1,0} \xrightarrow{\sigma_1} q_{1,1} \xrightarrow{\sigma_2} \cdots \xrightarrow{\sigma_i} q_{1,i}$\end{small} can be found in \( \mathcal{A}_1 \), but \( \mathcal{A}_2 \) lacks a corresponding transition at step \( i \). This contradicts the assumption that all finite paths \( \rho^{(n)} \) are simulable in \(\mathcal{A}_2 \).  

Case 2 (Infinite Rejection): 
If \( \mathcal{A}_2 \) can be driven by \( \sigma_1 \) fully, let \( q_{2,c} \) denote its state after \( \sigma_1 \). The infinite suffix \( \sigma_2^\omega \) induces a path in \( \mathcal{A}_2 \):  
\vspace{-0.3cm}

\[
\rho_2^{\omega}: q_{2,c} \xrightarrow{\sigma_2} q_{2,c}^{(1)} \xrightarrow{\sigma_2} q_{2,c}^{(2)} \xrightarrow{\sigma_2} \cdots,
\]  

forming an infinite sequence of states. As \( \sigma \notin L(\mathcal{A}_2) \), \( \rho_2^{\omega} \) must fail the Büchi condition. By the pigeonhole principle ( Jukna, 2011), there exists a strongly connected component (SCC) \( C \subseteq Q_2 \) visited infinitely often in \( \rho_2^{\omega} \), containing no accepting states. Let \( k \in \mathbb{N} \) be the minimal number such that \( \mathcal{A}_2 \)’s path on \( \sigma_2^k \) closes a cycle within \( C \). 
Then, the finite path \begin{small}
    $\quad q_{1,0} \xrightarrow{\sigma_1} q_{1,c} \xrightarrow{\sigma_2} \cdots \xrightarrow{\sigma_2} q_{1,c} \quad (\text{\( k+1 \) states})$\end{small} can be found in \( \mathcal{A}_1 \), but \( \mathcal{A}_2 \)’s corresponding path \( \rho_2^{w}\) either terminates prematurely or enters \( C \). Both violate the assumption that \( \mathcal{A}_2 \) simulates all finite paths.  

In either case, the existence of \( \rho \) which is simulable in $\mathcal{A}_1$ but not simulable in $\mathcal{A}_2$ is unavoidable, contradicting the assumption.  \hfill *
\end{proof}

\begin{proposition}\label{prop:time}
Let $\mathcal{A}_1 = (\Sigma, Q_1, \mathcal{I}_1, \mathcal{F}_1, \delta_1)$ and $\mathcal{A}_2 = (\Sigma, Q_2, \mathcal{I}_2, \mathcal{F}_2, \delta_2)$ be two Büchi automata. If a counterexample word \(\sigma \in \Sigma^\omega\) exists (as defined in Theorem~\ref{theo:counterexample}), then the counterexample path \(\rho\) (as described in Theorem~\ref{theo:counterexample}) can be identified in time
$O(n_1 \times n_2 \times m)$ in the worst-case,
where \(n_1 = |Q_1|\), \(n_2 = |Q_2|\) and $m=|\Sigma|$, where \(|\cdot|\) being the cardinality of a set.
\end{proposition}

\begin{proof}  
Define \(n_1 = |Q_1|\), \(n_2 = |Q_2|\), and let \(m= |\Sigma|\). The supergraph \(\mathcal{G}(V, E)\) as defined in Definition~\ref{def:supergraph} is constructed over state pairs \((q_1, q_2) \in Q_1 \times Q_2\), with:
$|V|\leq n_1 \times n_2$. We use the Breadth-first search(BFS) Algorithm to traverse from all state pairs at most once to detect a counterexample. Assume the size of the alphabet is fixed and finite, i.e., \(|\Sigma| = m\), then each pair requires exactly \(m\) comparisons. Therefore, the time complexity of transition checks $O(\cdot)$ is bounded by:
\(
O(\cdot)\leq O(|V| \times m) \leq O(n_1 \cdot n_2 \cdot m).
\)

When a mismatch is found, the counterexample path is constructed via backtracking from the visited state map (as also illustrated in Algorithm 1). The length of this path is at most the depth of the BFS tree, i.e., \(n_1 \times n_2\), and can be performed in linear time. Since path construction occurs once and after all comparisons, it does not affect the asymptotic bound.

Even if all vertices must be validated for acceptance conditions, the total operations remain dominated by \(|V|\). Thus, the counterexample trace is obtained in time:  
$O(n_1 \times n_2 \times m) $. \hfill *
\label{proof:time}
\end{proof}

\begin{algorithm}[h]
\caption{Comparison Path Storage}
\label{alg:TraceStore}
\textbf{Input}: Two automata $\mathcal{A}_1$, $\mathcal{A}_2$, maximum search depth $maxdepth$ \\
\textbf{Output}: A counterexample path string, or \texttt{null} if no counterexample is found
\begin{algorithmic}[1] 
\STATE Initialize $post[state][symbol]$ for state transitions
\STATE Initialize queue with initial state pairs $(state1, state2)$ from $\mathcal{A}_1$, $\mathcal{A}_2$
\STATE Initialize visited set to track explored state pairs
\WHILE{queue is not empty}
    \STATE Dequeue $(state1, state2)$
    \IF{$state1$ is in visited}
        \IF{$state2$ is in visited[$state1$]}
            \STATE \textbf{continue}
        \ENDIF
    \ENDIF
    \STATE Add $state2$ to visited[$state1$]
    \FOR{each symbol $s$}
        \STATE $nextStates1 \gets post[state1][s]$
        \STATE $nextStates2 \gets post[state2][s]$
        \IF{$nextStates1 \neq nextStates2$}
            \STATE Build and store the path from $(state1, state2)$
            \STATE \textbf{return} the path
        \ENDIF
    \ENDFOR
\ENDWHILE
\end{algorithmic}
\vspace{-0.1cm}
\end{algorithm}

To maximize the utility of counterexample paths, we introduce \emph{LLM-as-a-Critic}. We discovered that enabling one LLM to understand another LLM can overcome the LLM’s lack of knowledge in formal verification and format conversion. 
First, we let \emph{LLM-as-a-Critic} directly analyze the counterexample path and provide suggestions for modifying the input LTL specification. 
Then, we pass the output of \emph{LLM-as-a-Critic} to \emph{User LLM}, which has access to Environmental Information, to perform the actual LTL modifications. 
This process not only enhances the extraction of counterexample information but also allows for more effective modifications by incorporating Environmental Information (cf. Section~\ref{section4.2} for the ablation experiment).

To illustrate this process, we revisit the running example.\\
\textbf{Example (continued)}
First, we input the counterexample path into the \emph{LLM-as-a-Critic}.
The obtained violation analysis and correct guidance are then be fed to the User LLM to modify the LTL specification. Although the LTL still fails the Inclusion Check, after iterating the automated verification loop three times, the final LTL that adheres to the rules is obtained as follows:

\vspace{-0.4cm}
\begin{small}
\begin{align}
    \varphi = {} & X(\text{destinationLeftOnMapleStreet}) \land {} \nonumber\\
    & G\big(\text{location\_start} \to (F(\text{str\_200m}) \lor G(\neg \text{str\_200m}))\big) \land {} \nonumber\\
    & X\Big( G\big(\text{right\_turn\_Maple\_St} \to F(\text{straight\_500m} \;\land {} \nonumber\\
    & X(\text{left\_turn\_Oak\_St} \land F(\text{straight\_300m})))\big)\Big)\label{eq:ltl_balanced}
\end{align}
\end{small}
\vspace{-0.5cm}

Our framework prioritizes safety-compliance over strict semantic fidelity to the Desired Task: when the original route violates safety, it is reordered with a safe alternative that still reaches the same destination. In this example, the initial LTL formula already satisfies the latter part of the Safety Constraint, so the segment from “turn right onto Maple Street” remains unchanged. And because the Safety Constraint requires the journey to begin by proceeding straight or making a right turn, we add the \texttt{straight\_200m} restriction. It also stipulates that the destination is reached after a left turn followed by straight driving. Using Environmental Information, we identify that the last left turn occurs after entering Maple Street and therefore add $X(\text{destinationLeftOnMapleStreet})$ to ensure correct arrival. \hfill *

\textit{Remark 4.} Our ideal objective is to search for an LTL specification that satisfies all safety constraints while making the minimal semantic modification to the Desired Task. In practice, we set a maximum number of repair rounds and a timeout to prevent infinite correction loop for the Desired Task over which a safety-compliant case can't be found.

Once the Desired Task in LTL format passes the semantic check, a safety-compliant LTL $\varphi^{(k)}$ after $k$ rounds with respect to predefined safety constraint is obtained. Note that based on our framework's mechanism and the Inclusion Check's mathematical rigor, it is guaranteed that \emph{any LTL specification that is not safety-compliant will not be output}, which is critical for its involvement in subsequent applications. This would also be shown by the experiment in Section~\ref{section4.2}.

\section{Finetuning Approach}\label{sec:4}
\vspace{-0.1cm}
\subsection{Construction of Datasets}\label{subsec4.1}

A major limitation of existing NL2LTL datasets is their abstract APs lacking practical context. We emphasize human-centered APs to aid counterexample-driven correction. Since our goal is to fine-tune a lightweight LLM, the dataset must be compact yet effective.

We chose the traffic navigation as the primary scenario. To support the three-stage-fine-tuning process of our framework, we construct dedicated datasets for each stage, following a generate–filter–refine paradigm that ensures syntactic correctness, semantic alignment, and safety-compliance, respectively.

1) For stage-1, 50 navigation instruction templates were expanded via GPT-4 into NL-LTL pairs. Generated formulas were validated with SPOT ~\cite{33}, retaining only syntactically correct ones. We first perform a manual semantic check to ensure each natural language instruction accurately reflects the intent of its corresponding LTL formula. Then, we standardize and refine the wording of all instructions using a simplified prompt template, making their style consistent with downstream applications. The final result is a set of 200 high-quality Instruction-LTL training pairs. 

2) For stage-2, erroneous LTL specifications from generation were annotated with error types and paired with corrected versions passing SPOT, forming 225 Instruction–Corrected LTL pairs across 69 error types.

3) For stage-3, using RABIT~\cite{37}, we curated 50 triples (Instruction-Counterexample-Corrected LTL) from cases that initially failed Inclusion Checks. These rare and time-consuming cases make the collection especially valuable for training; we demonstrate its effectiveness in Section~\ref{subsec:finetuneproof}.

The Stage-1 dataset contains formulas with up to 8 nesting depths, with 99.0\% conjunctions and 39.5\% until operators. Detailed dataset statistics are provided in Table~\ref{tab:benchmark-stats}. For evaluation, we construct a set of 300 naturalistic navigation tasks featuring diverse logical structures and previously unseen combinations of atomic propositions. This dataset is completely held out from all training stages and serves as the evaluation benchmark for all LTL specification generation tasks in our work. 
\vspace{-0.2cm}
\begin{table}[!htbp]
\centering

\caption{Statistics of stage-1 dataset}
\label{tab:benchmark-stats}
\begin{tabular}{ll}
\hline
\textbf{Metric} & \textbf{Value / Statistics} \\ \hline
Formula Length (chars) & 55--267 (median 167) \\
No. of Unique APs & 3--15 (median 9) \\
No. of Temporal Operators & 2--10 (median 6) \\
Max Nesting Depth & 8 \\
\hline
\textit{Operator Frequency:} & \\
Conjunction ($\wedge$) & 99.0\% (198/200) \\
Disjunction ($\vee$) & 37.0\% (74/200) \\
Global ($G$) & 97.0\% (194/200) \\
Until ($U$) & 39.5\% (79/200) \\
\hline
\end{tabular}
\end{table}

\vspace{-0.1cm}
\subsection{Model Fine-tuning}
\vspace{-0.1cm}
We employ LLaMA-3-8B-Instruct as the edge-side \emph{User LLM}, fine-tuned using 4-bit QLoRA to ensure on-device efficiency. This parameter-efficient strategy enables the incremental injection of repair knowledge across stages while avoiding catastrophic drift. 
\vspace{-0.2cm}
\subsection{Proof of Efficiency}\label{subsec:finetuneproof}
\vspace{-0.1cm}
Ablation experiments evaluate the impact of each stage in our progressive fine-tuning strategy. Stage-1 focuses on improving the accuracy (i.e., the correctness of initial NL2LTL translation before any repair or refinement). Stage-2 enhances the model’s ability to generate syntactically correct formulas and efficiently self-correct. Stage-3 further equips the model to align with safety constraints and enhances its ability to perform semantic repair.

\textbf{Stage-1} Similar to ~\cite{NL2TL}, we adopt binary accuracy (i.e., 100\% correct or not) to measure raw translation performance, although this is not our primary focus. The evaluation is conducted on the 300-sample evaluation benchmark with 6 different settings. 

\begin{itemize}
\setlength\itemsep{-0.3pt}
\setlength\topsep{-0.3pt}
\setlength\parsep{0pt}
\setlength\partopsep{0pt}
  \item \textbf{LLaMA-Origin}: the base LLaMA-3-8B-Instruct model without fine-tuning.
  \item \textbf{GPT-4}: zero-shot prompting using the same task format.
  \item \textbf{LLaMA-200-raw}: LLaMA fine-tuned on 200 pairs filtered only by \texttt{SPOT} for syntactic validity (no semantic cleaning).
  \item \textbf{LLaMA-200-clean}: LLaMA fine-tuned on 200 pairs after both syntactic validity and manual semantic refinement.
  \item \textbf{LLaMA-35k-lifted}: LLaMA fine-tuned on 35,000 NL–LTL pairs adapted from the 28k lifted STL corpus in \cite{NL2TL}. Each STL formula is first converted into an LTL specification, where atomic propositions are rewritten to represent traffic navigation instructions. The corresponding natural language descriptions are then generated by GPT-4.
  \item \textbf{LLaMA-200-refined}: LLaMA fine-tuned on 200 cleaned pairs with deeper refinement and instruction-level optimization (cf.\ Section~\ref{subsec4.1}).
\end{itemize}
\vspace{-0.1cm}

The results (cf. Table~\ref{tab:combined-efficiency}) highlight the importance of data quality: simply cleaning a 200-example corpus (LLaMA-200-clean) narrows the gap to GPT-4, while a refined dataset makes LLaMa outperform GPT-4 with a lightweight 8B model. Conversely, a much larger but less targeted corpus (LLaMA-35k-lifted) degrades performance confirming that \emph{task-specific, semantically aligned data outweighs sheer quantity} for compact LLMs in structured specification generation. Among Stage-1 configurations, LLaMA-200-refined achieves the highest accuracy and is therefore selected for subsequent fine-tuning.

\textbf{Stage-2} To evaluate the effect of syntactic correction fine-tuning, we measure the \emph{average number of syntax correction iterations} required to produce a syntactically correct LTL formula. This metric reflects both the model’s initial correctness and its ability to efficiently repair errors when necessary.

As shown in Table~\ref{tab:combined-efficiency}, 
\texttt{AutoSafeLTL} achieves the best results, reducing the average to \textbf{0.56} versus 1.72 at Stage-1—a 67\% drop—outperforming LLaMA-Origin and GPT-4. This shows Stage-2 fine-tuning effectively instills syntactic knowledge, with many formulas correct on the first attempt.

\textbf{Stage-3} To evaluate the effect of semantic correction fine-tuning, we measure the \emph{average number of semantic correction iterations}—i.e., revisions required to achieve an \texttt{Included} status from a safety-violating LTL formula. This metric reflects the model’s \emph{intrinsic safety capability} and ability to efficiently repair non-compliant formulas.

The final stage further enhances semantic alignment through verification-guided correction. \texttt{AutoSafeLTL (Stage-3)} achieves the best performance on both syntax (\textbf{0.51} fixes) and semantic repair (\textbf{4.3} fixes), outperforming all baselines including GPT-4. And since Stage-3 introduces perturbations during counterexample-driven refinement, which may reduce surface-level matching but improve safety-compliance, its accuracy (93.6\%) is slightly below the Stage-1's peak, yet still outperforms GPT-4. This aligns with our main goal—ensuring safety-compliant LTLs.

\section{Framework Evaluation}\label{sec:eva}

\subsection{Setup}
We evaluate our framework's end-to-end effectiveness under a cloud–edge collaborative architecture, using \textit{violation counts}, i.e., safety violations detected via Inclusion Check against predefined safety constraints, as the primary metric. Lower violation counts indicate better safety-compliance.

We deploy \texttt{GPT-4} as cloud-edge Agent LLMs without fine-tuning, accessed via API and prompt engineering. Our evaluation is conducted on the same evaluation benchmark (cf. Section~\ref{subsec4.1}).

We compare our approach against several baselines:

\begin{itemize}\itemsep 0pt
    \item \textbf{LLaMA\&GPT-4}: zero-shot prompting using the same template as our method.
    \item \textbf{\texttt{nl2spec}}~\cite{16}: a LTL extraction tool. We adopt its default settings—\texttt{GPT-3.5-turbo} as the backend, \texttt{minimal} extraction mode, and \texttt{Number of tries} set to 3.
    \item \textbf{AutoSafeLTL}: our full pipeline under the collaborative cloud–edge setting.
\end{itemize}

\begin{table*}[h]
\centering

\caption{Violation rate and interaction rounds on the benchmark dataset. 
\textbf{B} denotes the Automated Verification component; 
\textbf{C}, the \emph{LLM-as-an-Aligner}; and 
\textbf{D}, the \emph{LLM-as-a-Critic}.}
\label{tab:ev}

\renewcommand{\arraystretch}{1.3} 
\begin{tabular*}{\textwidth}{@{\extracolsep{\fill}} l cccccc}
\toprule
 & \textbf{LLaMA+B} & \textbf{GPT-4+B} & \textbf{nl2spec+B} & \textbf{AutoSafeLTL} & \textbf{AutoSafeLTL-C} & \textbf{AutoSafeLTL-D} \\
\midrule
Violation (\%)     & 0    & 0    & 0    & 0    & 0    & 0 \\
Avg. Interaction   & 16.2 & 7.8  & 6.5  & \textbf{4.3} & 15.0 & 11.0 \\
\bottomrule
\end{tabular*}

\end{table*}

Notably, many real-world navigation instructions in our dataset lack a rigid logical structure, resulting in very low translation coverage by \texttt{nl2spec} (less than 5\% of original inputs produce valid LTL). For fair comparison, we use \texttt{GPT-4o} to rewrite all Desired Tasks into logically enhanced forms before \texttt{nl2spec} translation. These rephrased instructions serve as its evaluation input.

We also conduct ablation studies to investigate the necessity of both \emph{User LLM} and \emph{Agent LLM} in our framework. These variants isolate the contribution of each model and allow us to assess how collaboration between cloud and edge components influences final safety outcomes.

\begin{table}[h]
\vspace{-0.3cm}
\centering
\small
\caption{Stage-wise performance of our framework. Accuracy reflects the initial generated LTL's quality; Syn./Sem. Fixes indicate average iterations for syntax/semantic repair.}
\label{tab:combined-efficiency}
\begin{tabularx}{\linewidth}{lXXX}
\toprule
\textbf{Model} & \textbf{Accuracy (\%)} & \textbf{Syn. Fixes} & \textbf{Sem. Fixes} \\
\midrule
LLaMA-Origin            & 86.7 (260/300)            & 2.1   & 16.2 \\
GPT-4                   & 92.3 (277/300)            & 0.85  & 7.8  \\
AutoSafeLTL (Stage-1)   & \textbf{95.7 (287/300)}   & 1.72  & 7.1  \\
LLaMA-200-raw   & 91.3 (274/300)            & --    & --   \\
LLaMA-200-clean & 92.7 (278/300)            & --    & --   \\
LLaMA-35k-lifted& 90.0 (270/300)            & --    & --   \\
AutoSafeLTL (Stage-2)   & 94.3 (283/300)            & 0.56  & 6.8  \\
\textbf{AutoSafeLTL (Stage-3)} & 93.6 (281/300) & \textbf{0.51} & \textbf{4.3} \\
\bottomrule
\end{tabularx}

\end{table}
\vspace{-0.3cm}
\begin{table}[h]
\vspace{-0.3cm}
\centering
\small
\setlength{\tabcolsep}{4pt}
\caption{Initial violation rate comparison. \textbf{AutoSafeLTL*} uses only \emph{User LLM} without automated verification.}
\label{tab:initialviolation}
\begin{tabularx}{\linewidth}{
    >{\hsize=1.2\hsize}X
    >{\centering\arraybackslash\hsize=0.8\hsize}X
    >{\centering\arraybackslash\hsize=0.8\hsize}X
    >{\centering\arraybackslash\hsize=0.8\hsize}X
    >{\centering\arraybackslash\hsize=1.0\hsize}X}
\toprule
 & \textbf{llama} & \textbf{gpt-4} & \textbf{nl2spec} & \textbf{autosafeltl*} \\
\midrule
violation (\%) & 100 & 100 & 100 & \textbf{98} \\
\bottomrule
\end{tabularx}

\end{table}

\subsection{Quality of Initial Translation.}\label{section4.1}

We first evaluate the \emph{initial violation rate} on the benchmark—i.e., the proportion of initially generated LTL formulas that conflict with safety-compliance. Each initial LTL formula undergoes a single Inclusion Check: those marked \textsc{INCLUDED} are considered compliant, all others are violations.

Safety-compliance is often overlooked in prior LTL generation research. As shown in Table~\ref{tab:initialviolation}, prior approaches 
 exhibit a \textbf{100\%} violation rate, indicating that none of their generated LTL formulas initially conform to the safety constraints. In contrast, our \texttt{AutoSafeLTL*}—a variant that uses only the \emph{User LLM} without invoking the automated verification loop—achieves a violation rate of \textbf{98\%}, successfully producing safety-compliant LTL at generation time.

This result, though modest, is significant: it  confirms that progressive fine-tuning allows the model to internalize safety priors and generate compliant LTLs natively—a feat unmatched by baselines. It highlights the importance of training-level safety awareness for critical CPS applications.

\subsection{Quality of Semantic Faithfulness}\label{section4.3}

While our primary focus is provable safety-compliance rather than semantic faithfulness, we evaluated our framework's ability to preserve user intent. Following the protocol in \cite{NL2TL}, we tested models from all three fine-tuning stages on an extended dataset of 1,000 navigation tasks, using their consistency metric to verify that core objectives (e.g., waypoints, actions) remain intact in the generated LTL.

The semantic consistency scores across the three fine-tuning stages are 94.7\% (S1), 92.5\% (S2), and 91.1\% (S3). A slight decline is observed across the stages, which aligns with our design goal: later stages increasingly prioritize safety-compliant repair when conflicts with safety constraints arise. 
However, the final model (Stage-3) still maintains a high consistency score ($>91\%$), demonstrating that AutoSafeLTL effectively minimizes semantic drift while enforcing rigorous safety guarantees.
\vspace{-0.1cm}

\subsection{Quality of Verified Translation and Ablation Study}\label{section4.2}

Next, we evaluate the effectiveness of our safety assurance component \emph{Automated Verification module}. \textit{Safety-compliant} is defined by an LTL formula passing the Inclusion Check; we also measure the iterations required to reach this state as an efficiency metric.


Applying our full verification pipeline (\texttt{+B}) to LTLs generated by multiple baselines like \texttt{LLaMA}, \texttt{GPT-4}, and \texttt{nl2spec} yields a 0\% violation rate across all models (Table~\ref{tab:ev}), confirming its generalizability. Notably, \texttt{AutoSafeLTL} achieves this with the fewest interaction rounds (5.7)—significantly outperforming \texttt{GPT-4+B} (7.8) and \texttt{LLaMA+B} (16.2)—demonstrating superior efficiency in aligning with safety constraints.

To further understand the role of each cloud-side agent, we perform ablation experiments by selectively removing \emph{LLM-as-an-Aligner} (\texttt{AutoSafeLTL-C}) and \emph{LLM-as-a-Critic} (\texttt{AutoSafeLTL-D}). Both variants still achieve full compliance, but at the cost of more iterations (15.0 and 11.0, respectively), highlighting the complementary importance of both agents in minimizing verification overhead.


\subsection{Inference Latency and Efficiency}\label{section4.4}
To assess the practical feasibility of \texttt{AutoSafeLTL} for real-time applications, we conducted a latency analysis on an NVIDIA A6000 GPU. With a batch size of 20 and five parallel safety constraints, the framework averages 0.27s per iteration. Given the 4.3-round average to reach full compliance (Table~\ref{tab:ev}), the total end-to-end latency is approximately 1.16s, making it compatible with high-level planning cycles. Detailed latency statistics are provided in Tabel~\ref{tab:latency}. Notably, formal verification (BA Construction and Inclusion Check) accounts for only $\approx$ 4\% of total time. This confirms that our counterexample path extraction algorithm is highly efficient and that the computational bottleneck lies in LLM inference rather than the safety assurance mechanism.

\begin{table}[h]
\centering
\small
\setlength{\tabcolsep}{4pt}
\caption{Component Latency Breakdown (per iteration round).}
\label{tab:latency}
\begin{tabularx}{\linewidth}{
    >{\hsize=1.4\hsize}X
    >{\centering\arraybackslash\hsize=0.8\hsize}X
    >{\centering\arraybackslash\hsize=0.8\hsize}X}
\toprule
\textbf{Component} & \textbf{Time (s)} & \textbf{Share (\%)} \\
\midrule
LLM Generation & 0.7600 & 95.9\% \\
BA Construction & 0.0097 & 1.2\% \\
Safety Verification & 0.0227 & 2.9\% \\
\textbf{Total (Single-stream)} & \textbf{0.7924} & \textbf{100\%} \\
\bottomrule
\end{tabularx}
\end{table}

\section*{Acknowledgment}
This work was supported by Guangzhou-HKUST(GZ) Joint Funding Program (Grant No.~2025A03J4493), Education Bureau of Guangzhou Municipality, Guangdong Provincial Project 2024QN11X053, and the Youth S\&T Talent Support Programme of GDSTA (SKXRC2025468).

\vspace{-0.2cm}
\bibliographystyle{IEEEtran}
\bibliography{ref}

\end{document}